# COMMENTS ON THE PAPER "FUNDAMENTAL STELLAR PHOTOMETRY FOR STANDARDS OF SPECTRAL TYPE ON THE REVISED SYSTEM OF THE YERKES SPECTRAL ATLAS "

## BY H.L. JOHNSON AND W.W. MORGAN (1953)

SIDNEY VAN DEN BERGH

Dominion Astrophysical Observatory
Herzberg Institute of Astrophysics
National Research Council of Canada
5071 West Saanich Road
Victoria, British Columbia, V8X 4M6
Canada

Photometry and spectroscopy are among the most important and fundamental types of astronomical observations. By the middle of the twentieth century most spectral classification was defined in terms of MKK type standards (Morgan, Keenan & Kellerman 1943). However, for a variety of reasons (Weaver 1947), the establishment of photometric standards remained problematic. Commission 25 (Photometry) of the International Astronomical Union devoted seemingly endless years to discussions on the establishment of an international photometric standard system. Weaver (1952) advocated that a new sub-commission should be created which would have as its task "The establishment of an 'ideal' photo-electric color and magnitude system." A subsequent IAU resolution (Oosterhoff 1954) suggested that "La Commission recommende que soit constituée une nouvelle sous-commission des étalons de magnitudes stellaire....". Additional problems were: (1) The Commission could not decide between monochromatic magnitudes and broad-band magnitudes, and (2) some advocated a two-color system (Hertzsprung 1950), whereas others (Becker 1946) favored a three-color (photographic, photovisual, red) photometric system. In his characteristically energetic fashion Johnson (Johnson & Morgan 1951) then took the bull by the horns and started to observe so many stars and clusters on his UBV system that it soon became the <u>de facto</u> international photometric system[1].

---

[1]     In his autobiography Morgan (1988) writes "The UBV system of magnitudes and colors was invented by me during the period 1947-51".

---

The UBV system was defined in terms of a set of Schott/Corning glass filters, in front of a 1P21 photomultiplier, cooled by dry ice. Particularly important features of the system are: (1) it includes a U filter, which proves to be especially useful for the study of hot stars and of metallicity effects, and (2) the short-wavelength cut-off is defined by the transmission function of the filters and the spectral sensitivity of the 1P21, rather than by the atmosphere. However, the fact that Johnson's 1P21 had an above-average red sensitivity causes others, who want to transform their data to the UBV system, some problems (Gehrels 1998).

Johnson's photometry and Morgan's spectroscopy were combined in the paper (Johnson & Morgan 1953), which is the subject of the present essay. The MK system presented in that paper provided minor improvements on the older MKK system of spectral classification. However, the UBV photometric system was a major improvement over previous photometric systems. It soon became clear that combining UBV photometry and MK spectral classification provided a powerful technique for the exploitation of Baade's (1944) new ideas on stellar populations. For early-type stars of Baade's Population I, UBV photometry (or BV photometry and MK spectroscopy) enabled one to determine the reddening of individual stars. Furthermore calibration of the MK luminosity classes, and main sequence fitting of distant clusters to nearby clusters of known distance, made it possible to determine distances to remote clusters. Such techniques, in conjunction with luminosity calibration of MK spectral classes, allowed Morgan, Whitford & Code (1953) to show convincingly that the galaxy is a two-arm spiral galaxy. Furthermore Burbidge & Sandage (1958) used photometry on the UBV system to show how the color-magnitude diagrams of open clusters evolve with age. Among nearby unreddened stars UBV photometry made it possible to determine the ultraviolet excess, and hence the metallicities, of individual stars. Such observations provided the vital underpinning of the work by Eggen, Lynden-Bell & Sandage (1962), which established the first modern paradigm for the formation of our Milky Way system. Later UBV photometry of individual stars in Population II systems, such as M 3, M 13, M 15, and M 92 (Sandage 1969, 1970) showed that such observations could provide a great deal of information on the ages and metallicities of individual globular clusters.

Primarily because of its importance to studies of stellar populations, the paper by Johnson & Morgan (1953) soon became the most cited paper in the astronomical literature. This popularity is also attested to by the fact that the binding of our Observatory's copy of ApJ 117 is broken, so that individual pages of the Johnson and Morgan paper had to be re-attached with tape.

It is a pleasure to thank Helmut Abt, Tom Gehrels, Arlo Landolt and an anonymous referee for their historical recollections.